\documentclass{ws-ijmpe}

\usepackage{graphicx} 

\begin{document}

\markboth{FRANCESCO D'ERAMO, HONG LIU, KRISHNA RAJAGOPAL}{Jet Quenching Parameter via Soft Collinear Effective Theory (SCET)}

\catchline{}{}{}{}{}

\title{Jet Quenching Parameter via Soft Collinear Effective Theory (SCET)}

\author{FRANCESCO D'ERAMO, HONG LIU, KRISHNA RAJAGOPAL}

\address{Center for Theoretical Physics, Massachusetts Institute of Technology \\
Cambridge, MA 02139, United States \\
fderamo@mit.edu
}



\maketitle


\begin{history}
\received{(received date)}
\revised{(revised date)}
\end{history}

\begin{abstract}
We analyze the transverse momentum broadening in the absence of radiation of an energetic parton propagating through quark-gluon plasma via Soft Collinear Effective Theory (SCET).  We show that the probability for picking up transverse momentum $k_\perp$ is given by the Fourier transform of the expectation value of two transversely separated light-like path-ordered Wilson lines. The subtleties about the ordering of operators do not change the $\hat q$ value for the strongly coupled plasma of $\mathcal{N}=4$ SYM theory.
\end{abstract}

\section{Introduction}
One of the central discoveries made in experimental heavy ion collisions at the Relativistic Heavy Ion Collider (RHIC) at Brookhaven National Laboratory is that the droplets of quark-gluon plasma produced in these collisions are sufficiently strongly coupled that they are able to ``quench jets''~\cite{Arsene:2003yk}. 
In the high energy limit the parton loses energy dominantly by inelastic processes that are the QCD analogue of bremsstrahlung~\cite{Gyulassy:1993hr,Baier:1996sk}. It is crucial to the calculation of this radiative energy loss process that the incident hard parton, the outgoing parton, and the radiated gluons are all continually being jostled by the medium in which they find themselves: they are {\it all} subject to transverse momentum broadening.  

The transverse momentum broadening of a hard parton is described by $P(k_\perp)$\footnote{We normalize $P(k_\perp)$ as $\int \frac{d^2 k_{\perp}}{(2\pi)^2} P (k_{\perp}) = 1$ . }, defined as the probability that after propagating through the medium for a distance $L$ the hard parton has acquired transverse momentum $k_\perp$.\footnote{$k_\perp$ is the two-dimensional vector $\vec{k}_\perp$ in transverse momentum space. We will also represent two-dimensional vectors in transverse coordinate space by $x_\perp$, again without the vector symbols.} The mean transverse momentum picked up by the hard parton per unit distance travelled results in:
\begin{equation}
\hat q \equiv \frac{\langle k_\perp^2 \rangle}{L} =
\frac{1}{L} \int \frac{d^2 k_{\perp}}{(2\pi)^2} k_\perp^2 P (k_{\perp})\ .
\label{qhatFirstTime}
\end{equation}
The quantity $\hat q$ is called the ``jet quenching parameter'' because it plays a central role in calculations of radiative parton energy loss~\cite{Baier:1996sk,Wiedemann:2000za}, although it is {\it defined} via transverse momentum broadening only.  Radiation and energy loss do not arise in its definition, although they are central to its importance.

The calculation of parton energy loss and transverse momentum broadening involves widely separated scales:
\begin{equation}
Q \gg l_\perp \gg T,
\end{equation}
where $Q$ is the energy of the hard parton, $l_\perp$ the momentum of the radiated gluon transverse to that of the incident parton and $T$ the soft scale characteristic of the medium. We can ultimately hope for a factorized description, with physics at each of these scales cleanly separated at lowest nontrivial order in a combined expansion in the small ratio between these scales and in the QCD coupling $\alpha$ evaluated at scales which become large in the high parton energy limit.  And, most importantly, we can aspire to having a formalism in which corrections to this factorization are calculable systematically, order by order in these expansions.  No current theoretical formulation of jet quenching calculations is manifestly systematically improvable in this sense.  We take a small step toward such a description~\cite{D'Eramo:2010ak}: we formulate the calculation of transverse momentum broadening and the jet quenching parameter in the language of Soft Collinear Effective Theory (SCET)~\cite{Bauer:2000ew}, which has rendered the calculation of many other processes involving soft and collinear degrees of freedom systematically improvable. The problem that we analyze really does not use the full power of SCET, and the bigger payoff from a SCET calculation will come once radiation is included and once the analysis is pushed beyond the present leading order calculation.

\section{Set-up and relevance of ``Glauber gluons''}

Consider an on-shell high energy parton with initial four momentum\footnote{The light cone coordinates are defined by $q^{\pm} = \frac{1}{2} (q^0 \pm q^3)$.
}
\begin{equation}
q_0 \equiv (q_0^+, q_0^-, q_{0 \perp}) = (0, Q, 0)
\end{equation}
propagating through some form of QCD matter, which for definiteness we will take to be quark-gluon plasma (QGP) in equilibrium at temperature $T$ although the discussion of this paper would also apply to propagation through other forms of matter. We will assume throughout that $Q$ is very much larger than the highest momentum scales that characterize the medium, which for the case of QGP means 
$Q\gg T$.  Thus, we have a small dimensionless ratio in our problem:
\begin{equation} \label{hilim}
\lambda \equiv \frac{T}{Q} \ll 1 \ .
\end{equation}

\begin{figure}
\begin{center}
$\begin{array}{ccc}
\epsfxsize=2.2in
\epsffile{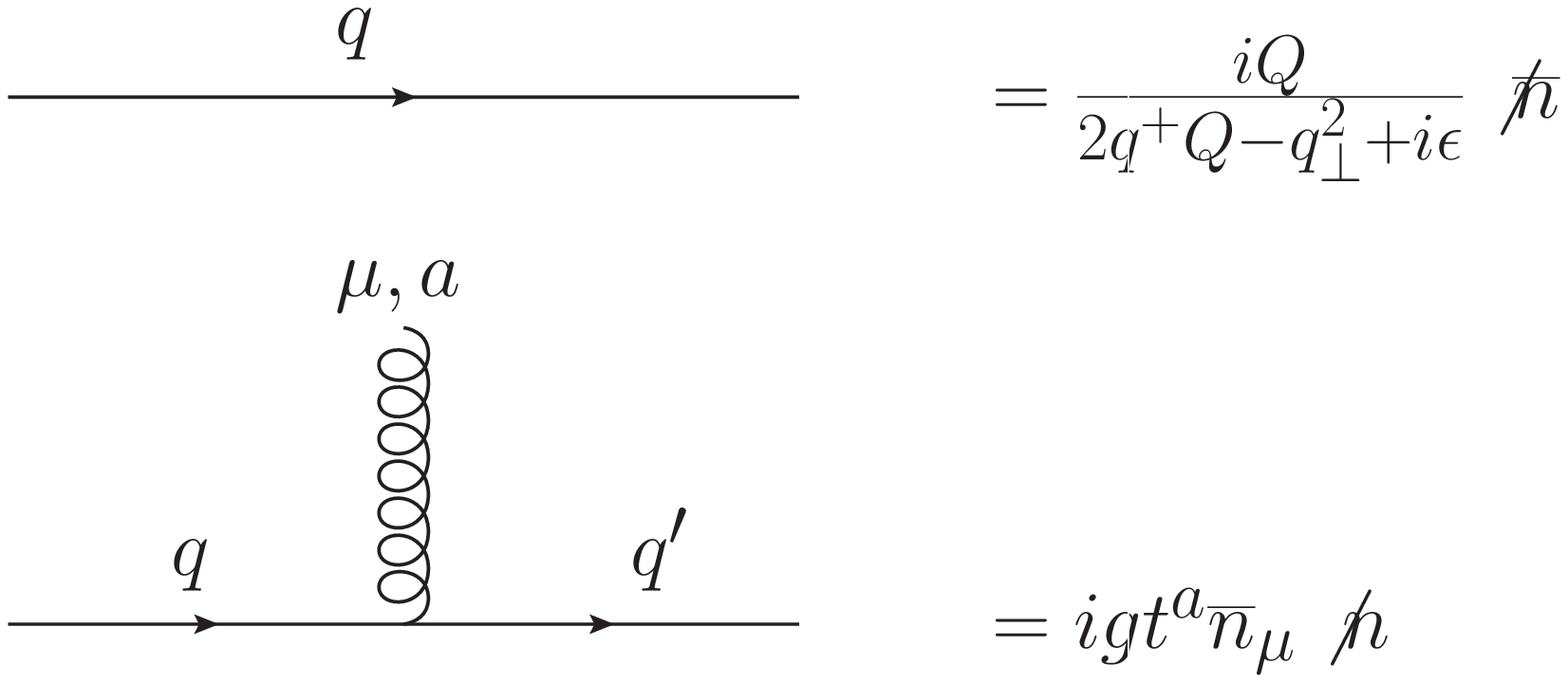} & \hspace{0.2cm} &
	\epsfxsize=2.5in
	\epsffile{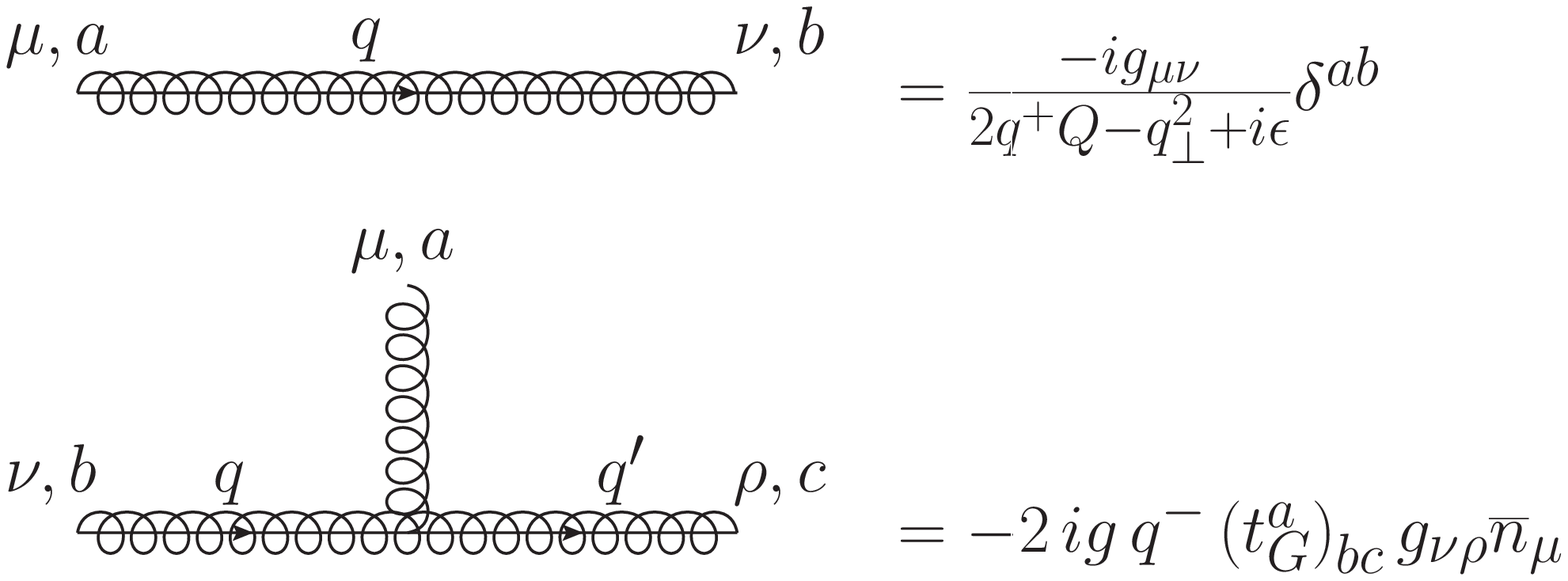} \\ [0.1cm]
\mbox{\bf (a)} & & \mbox{\bf (b)}
\end{array}$
\end{center}
\caption{Feynman rules for collinear partons interacting with Glauber gluons: (a) collinear quarks; (b) collinear gluons in Feynman gauge ($\alpha = 1$). The light cone vectors are defined as $\bar{n}\equiv \frac{1}{\sqrt{2}}\left(1,0,0,-1\right)$ and $n\equiv\frac{1}{\sqrt{2}}\left(1,0,0,1\right)$.}
\label{fig:Feynm}
\end{figure}

In the high energy limit the momentum broadening is induced by the interaction between the hard parton and gluons from the medium with momenta
 \begin{equation} \label{doM}
p \sim Q (\lambda^2, \lambda^2, \lambda)\ ,
\end{equation}
as first established by Idilbi and Majumder~\cite{Idilbi:2008vm}, who made the first attempt to extend SCET to describe hard jets in a dense medium.  Gluons in this kinematic regime are conventionally called ``Glauber gluons''. After absorbing or emitting Glauber gluons, the momentum of the hard parton is of order $q \sim Q (\lambda^2, 1, \lambda)$. We shall refer to modes with momenta of this parametric order as ``collinear''. The parton is only off-shell by of order $\lambda^2 Q^2\sim T^2$, and further absorption or emission of Glauber gluons keeps the parton off-shell by of the same order.\footnote{As does 
absorption or emission of "ultrasoft" modes with momenta $\sim Q(\lambda^2,\lambda^2,\lambda^2)$. They are simply a subset of the Glauber modes and thus are included in our analysis.  So too are all modes whose momenta are proportional to even higher powers of $\lambda$.  Interestingly this is true also for modes with momenta $\sim Q(\lambda^2,\lambda,\lambda)$, and our analysis of Glauber gluons applies to them also.} And yet, repeated absorption and emission of Glauber gluons continually kicks the hard parton and can result in significant transverse momentum broadening. The interaction vertex of each Glauber gluon with the parton is governed by $\alpha_s(T)$ and so can be strongly coupled. At a heuristic level, one can imagine Glauber gluons as a gluon background
surrounding the parton and as a result of frequent small kicks from this background, the parton will undergo Brownian motion in momentum space. The effective Lagrangian governing the interaction between collinear partons and Glauber gluons is derived using SCET in Refs.~\cite{D'Eramo:2010ak,Idilbi:2008vm}. The Feynman rules are given in Fig.~\ref{fig:Feynm}.

\section{Optical theorem and momentum broadening}
Our goal is to compute the probability distribution $P (k_{\perp})$ defined in the introduction, and the field theory tools we use to perform this computation are the $S$ matrix and the optical theorem. 

We shall imagine a cubic box with sides of length $L$ that is filled with the medium, and that satisfies periodic boundary conditions, leading to the quantization of momenta. We denote the single-particle in state describing the incident hard parton and the final state by $|\alpha \rangle$ and $|\beta \rangle$, respectively.\footnote{The box states have unit norm.} Since we are ignoring radiation, the only out states that we need consider are also single-particle states. 

The $S$-matrix element $S_{\beta\alpha}$ is defined as the probability amplitude for the process $\alpha \rightarrow \beta$. As usual, we first isolate the identity part of the S-matrix, $S_{\beta \alpha} = \delta_{\beta\alpha} + i M_{\beta\alpha}$, in so doing defining the interaction matrix element $M_{\beta\alpha}$. Conservation of probability implies unitarity for the $S$-matrix:
\begin{equation}
2 \,\Im\, M_{\alpha\alpha} = \sum_{\beta} |M_{\beta\alpha}|^2\ .
\label{eq:imp1}
\end{equation}
At this formal level, (\ref{eq:imp1}) would still be valid if we were including the effects of radiation, meaning that final states $|\beta\rangle$ would include many particle states. We are interested in computing the probability for the process $\alpha \rightarrow \beta$, where both states describe single particles and the final state $\beta$ differs from the initial state $\alpha$ only in its value of $k_{\perp}$. No other quantum numbers change. In particular, $\beta=\alpha$ corresponds to $k_{\perp}=0$. From the box normalization\footnote{As a consequence of the box normalization we have $\sum_{\beta} = L^2 \int \frac{d^2 k_{\perp}}{(2\pi)^2}$.} we obtain
\begin{equation}
P(k_{\perp}) = L^2 \left\{ \begin{array}{lc}
|M_{\beta\alpha}|^2 &\quad \beta \neq \alpha \\
1- 2\, \Im M_{\alpha\alpha} + |M_{\alpha\alpha}|^2 &\quad \beta=\alpha \ .
\end{array} \right.
\label{eq:PvsS}
\end{equation}
We first compute twice the imaginary part of the forward scattering amplitude, $2\,\Im M_{\alpha\alpha}$, by cutting the appropriate diagrams. Once we know $2\, \Im M_{\alpha\alpha}$ we can use the unitarity relation (\ref{eq:imp1}) to read off $\sum_\beta |M_{\beta\alpha}|^2$. Knowing the box normalization, we will immediately be able to identify $P(k_\perp)$.

\begin{figure}
\begin{center}
\includegraphics[scale=0.28]{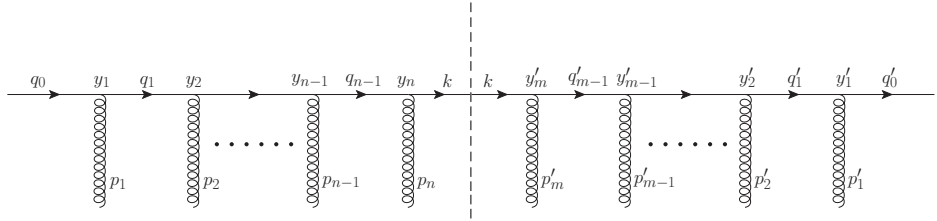}
\end{center}
\caption{Leading diagram for the transverse momentum broadening of a hard (collinear) parton, in the absence of any radiation. All the gluons are Glauber gluons.}
\label{fig:forw}
\end{figure}
The leading diagrams for the forward scattering amplitude are given in Fig.~\ref{fig:forw}, with the dashed line indicating where we cut the diagram and with all gluon lines being Glauber gluons. We will essentially consider the Glauber gluon insertions as external background fields and average them over the thermal ensemble only at the end of the calculation. That is, we first consider the propagation of the hard parton in the presence of one background field configuration, analyzing this problem including arbitrarily many interactions with the background field.  The nonperturbative physics of the medium does not enter this calculation. We shall then stop, leaving unevaluated in our answer the average over a thermal ensemble. 

We evaluate the forward scattering amplitude in the $Q\rightarrow\infty$ limit. We make the statement of this limit precise
\begin{equation}
 \label{PreciseLimit}
 Q \gg k_\perp^2 L \sim \hat q L^2\ .
\end{equation}
The physical significance of this criterion is that in the regime (\ref{PreciseLimit}) the distance $L$ that the hard parton propagates through the medium is short enough that the trajectory of the hard parton in position space remains well-approximated as a straight line, even though it picks up transverse momentum.  

The final result for the probability distribution $P(k_\perp)$ is
\begin{equation} \label{provb}
P(k_\perp) = \int d^2 x_{\perp} \, e^{-i k_{\perp} \cdot x_{\perp}}\,
 \mathcal{W}_{\mathcal{R}} (x_\perp) \ ,
\end{equation}
where $\mathcal{R}$ is the $SU(N)$ representation to which the collinear particle belongs and $\mathcal{W}_{\mathcal{R}} (x_\perp)$ is the expectation value of two light-like Wilson lines separated in the transverse plane by the vector $x_\perp$. Note that $\mathcal{W}_{\mathcal{R}} (x_\perp)$ depends only on the properties of the medium.  It is independent of the energy of the hard parton, meaning that so are $P(k_\perp)$ and $\hat q$ in the limit in which this hard parton energy is taken to infinity.  
Transverse momentum broadening without radiation thus does ``measure'' a field-theoretically well-defined property of the strongly coupled medium.  This is the kind of factorization that we hope to find in a systematically improvable calculation once radiation is included. Eq.~\eqref{provb} has been obtained previously by Casalderrey-Solana and Salgado and by Liang, Wang and Zhou using different methods~\cite{CasalderreySolana:2007zz}.

\section{$\hat q$ evaluation in $\mathcal{N}=4$ SYM theory revisited}
With $P(k_\perp)$ in hand, it is easy to obtain the jet quenching parameter $\hat q$
\begin{equation}
\hat{q} = \frac{\sqrt{2}}{L^-}  \, \int \frac{d^2 k_{\perp}}{(2\pi)^2}\,k_{\perp}^2 \, \int d^2 x_{\perp} \, e^{-i k_{\perp} \cdot x_{\perp}}\,
\mathcal{W}_{\mathcal{R}} (x_\perp) \ .
\label{qhatFromW}
\end{equation}
Crucially, and as a consequence of our field-theoretical formulation, we find that the ordering of operators in the expectation value $\mathcal{W}_{\mathcal{R}} (x_\perp)$ is {\it not} that of a standard Wilson loop --- the operators (like the color matrices)  are path ordered, whereas in a standard Wilson loop operators are time ordered. Because the operators are path ordered, the expectation value should be described using the Schwinger-Keldysh contour with one of the light-like Wilson lines on the ${\rm Im}\, t=0$ segment of the contour and the other light-like Wilson line on the ${\rm Im}\, t=-i\epsilon$ segment of the contour. This subtlety throws into question the calculation of $\hat q$ in the strongly coupled plasma of large-$N_c$ ${\cal N}=4$ supersymmetric Yang-Mills (SYM) theory reported in  Refs.~\cite{Liu:2006ug}.

We have to revisit the strong coupling calculation, and our procedure~\cite{D'Eramo:2010ak} is a specific example of the more general discussion of Lorentzian AdS/CFT given recently by Skenderis and Van Rees in Refs.~\cite{Skenderis:2008dh}, which we have followed. The subtleties introduced in the gravity calculation, corresponding to the subtlety of operator ordering in the field theory, turn out not to change the results of Refs.~\cite{Liu:2006ug} for $\mathcal{W}_{\mathcal{R}} (x_\perp)$ and $\hat q$ in ${\cal N}=4$ SYM theory. In fact, they solidify these results since it now turns out that there is  only one extremized string world sheet that is bounded by the light-like Wilson lines, and it is the one identified on physical grounds in Refs.~\cite{Liu:2006ug}. We quote the result
\begin{equation}
\hat q=\frac{\pi^{3/2}\Gamma(\frac{3}{4})}{\Gamma(\frac{5}{4})}\sqrt{\lambda} T^3 \ .
\label{qhatResult}
\end{equation}
This turns out to be in the same ballpark as the values of $\hat q$ inferred from RHIC data on the suppression of high momentum partons in heavy ion collisions~\cite{Liu:2006ug}.

\section*{Acknowledgements}
The work of HL was supported in part by a DOE Outstanding Junior Investigator grant. This research was supported in part by the DOE Offices of Nuclear and High Energy Physics under grants \#DE-FG02-94ER40818 and  \#DE-FG02-05ER41360..


\begin{thebibliography}{0}
\bibitem{Arsene:2003yk}
  I.~Arsene {\it et al.}  [BRAHMS Collaboration],
  Phys.\ Rev.\ Lett.\  {\bf 91}, 072305 (2003)
  [arXiv:nucl-ex/0307003];
  S.~S.~Adler {\it et al.}  [PHENIX Collaboration],
  Phys.\ Rev.\ Lett.\  {\bf 91}, 072303 (2003)
  [arXiv:nucl-ex/0306021];
  B.~B.~Back {\it et al.}  [PHOBOS Collaboration],
  Phys.\ Rev.\ Lett.\  {\bf 91}, 072302 (2003)
  [arXiv:nucl-ex/0306025];
  J.~Adams {\it et al.}  [STAR Collaboration],
  Phys.\ Rev.\ Lett.\  {\bf 91}, 072304 (2003)
  [arXiv:nucl-ex/0306024].
  
\bibitem{Gyulassy:1993hr}
  M.~Gyulassy and X.~n.~Wang,
  Nucl.\ Phys.\  B {\bf 420}, 583 (1994)
  [arXiv:nucl-th/9306003].

\bibitem{Baier:1996sk}
  R.~Baier, Y.~L.~Dokshitzer, A.~H.~Mueller, S.~Peigne and D.~Schiff,
  Nucl.\ Phys.\  B {\bf 484}, 265 (1997)
  [arXiv:hep-ph/9608322];
  B.~G.~Zakharov,
  JETP Lett.\  {\bf 65}, 615 (1997)
  [arXiv:hep-ph/9704255].
  
\bibitem{Wiedemann:2000za}
  U.~A.~Wiedemann,
  Nucl.\ Phys.\  B {\bf 588}, 303 (2000)
  [arXiv:hep-ph/0005129];
  M.~Gyulassy, P.~Levai and I.~Vitev,
  Nucl.\ Phys.\  B {\bf 594}, 371 (2001)
  [arXiv:nucl-th/0006010];
  X.~f.~Guo and X.~N.~Wang,
  Phys.\ Rev.\ Lett.\  {\bf 85}, 3591 (2000)
  [arXiv:hep-ph/0005044];
  X.~N.~Wang and X.~f.~Guo,
  Nucl.\ Phys.\  A {\bf 696}, 788 (2001)
  [arXiv:hep-ph/0102230];
  P.~B.~Arnold, G.~D.~Moore and L.~G.~Yaffe,
  JHEP {\bf 0206}, 030 (2002)
  [arXiv:hep-ph/0204343].
  
\bibitem{D'Eramo:2010ak}
 F.~D'Eramo, H.~Liu, K.~Rajagopal,
 [arXiv:1006.1367 [hep-ph]].
 
\bibitem{Bauer:2000ew}
 C.~W.~Bauer, S.~Fleming and M.~E.~Luke,
 Phys.\ Rev.\  D {\bf 63}, 014006 (2000)
 [arXiv:hep-ph/0005275];
%
 C.~W.~Bauer, S.~Fleming, D.~Pirjol and I.~W.~Stewart,
 Phys.\ Rev.\  D {\bf 63}, 114020 (2001)
 [arXiv:hep-ph/0011336];
%
 C.~W.~Bauer and I.~W.~Stewart,
 Phys.\ Lett.\  B {\bf 516}, 134 (2001)
 [arXiv:hep-ph/0107001];
%
 C.~W.~Bauer, D.~Pirjol and I.~W.~Stewart,
 Phys.\ Rev.\  D {\bf 65}, 054022 (2002)
 [arXiv:hep-ph/0109045].

\bibitem{Idilbi:2008vm}
 A.~Idilbi and A.~Majumder,
 Phys.\ Rev.\  D {\bf 80}, 054022 (2009)
 [arXiv:0808.1087 [hep-ph]].

\bibitem{CasalderreySolana:2007zz}
  J.~Casalderrey-Solana and C.~A.~Salgado,
  Acta Phys.\ Polon.\  B {\bf 38}, 3731 (2007)
  [arXiv:0712.3443 [hep-ph]];
  Z.~T.~Liang, X.~N.~Wang and J.~Zhou,
  Phys.\ Rev.\  D {\bf 77}, 125010 (2008)
  [arXiv:0801.0434 [hep-ph]].

\bibitem{Liu:2006ug}
  H.~Liu, K.~Rajagopal and U.~A.~Wiedemann,
  Phys.\ Rev.\ Lett.\  {\bf 97}, 182301 (2006)
  [arXiv:hep-ph/0605178];
  H.~Liu, K.~Rajagopal and U.~A.~Wiedemann,
  JHEP {\bf 0703}, 066 (2007)
  [arXiv:hep-ph/0612168].

\bibitem{Skenderis:2008dh}
  K.~Skenderis and B.~C.~van Rees,
  Phys.\ Rev.\ Lett.\  {\bf 101}, 081601 (2008)
  [arXiv:0805.0150 [hep-th]];
  K.~Skenderis and B.~C.~van Rees,
  JHEP {\bf 0905}, 085 (2009)
  [arXiv:0812.2909 [hep-th]];
  B.~C.~van Rees,
  Nucl.\ Phys.\ Proc.\ Suppl.\  {\bf 192-193}, 193 (2009)
  [arXiv:0902.4010 [hep-th]].

\end{thebibliography}
\end{document}